\documentclass[a4paper, conference]{IEEEtran}
\IEEEoverridecommandlockouts
\usepackage[nocompress]{cite}
\usepackage{amsmath,amssymb,amsfonts}
\usepackage{algorithmic}
\usepackage{graphicx}
\usepackage{textcomp}
\usepackage{xcolor}
\usepackage{amssymb}  
\usepackage{algorithm}
\begin{document}
    \title{\huge Study of Iterative Detection, Decoding and Channel Estimation for RIS-Aided MIMO Networks\vspace{-0.5em} }

\author{Roberto C. G. Porto and Rodrigo C. de Lamare \vspace{-1.1em}

\thanks{The authors are with the Centre for Telecommunications Studies, Department of Electrical Engineering, Pontifical Catholic University of Rio de Janeiro. Emails: camara@aluno.puc-rio.br, delamare@puc-rio.br}}

\maketitle

    \begin{abstract}
This work proposes an iterative detection, decoding and channel estimation scheme for multiple-antenna systems assisted by multiple reflective intelligent surfaces (RIS). A novel channel estimation technique that exploits low-density parity-check (LDPC) codes and iterative processing is developed to enhance estimation accuracy while reducing the number of required pilot symbols. The key idea is to exploit encoded pilots to improve the iterative process, enabling the use of not only pilot bits but also parity bits from the coded packet to refine channel estimation. Simulations analyze a sub-6 GHz scenario where the channel propagation is not sparse and multiple RIS are deployed, considering both LOS and NLOS conditions. Numerical results show significant performance gains for the proposed scheme and estimator over competing approaches. \vspace{0.25em}
\end{abstract}


    \vspace{-1em}
\section{Introduction}

Reconfigurable Intelligent Surfaces (RIS) have garnered significant attention as transformative technologies for the sixth generation (6G) of wireless communication networks. Unlike traditional approaches that rely on active signal transmission, RIS passively reshape the wireless environment by adjusting the reflection properties of electromagnetic waves \cite{9140329}. However, this capability comes at the cost of estimating a large number of channel coefficients \cite{10818440}, a task that is particularly challenging when dealing with multiple RIS. In such scenarios, the system effectively experiences an increased number of channels, further complicating the channel estimation with a limited number of pilot signals.

Several studies have investigated channel estimation techniques for RIS-assisted systems. In \cite{10818463}, a deep learning-based approach was proposed and evaluated using a compressed sensing method. Similarly, \cite{9127834} studied a deep neural network to improve compressive channel estimation by applying a complex-valued denoising convolutional neural network in millimeter-wave systems. Meanwhile, \cite{10614235} employed Bayesian learning for channel estimation in massive MIMO systems. A key insight in most channel estimation studies is that the RIS reflect signals from all users to the BS over the same channels. This results in correlation among the user-RIS-BS reflected channels, which can be exploited to reduce channel estimation overhead. In \cite{9130088}, a three-phase framework is proposed to exploit this correlation, where the first phase assumes that the RIS can be turned off, effectively nulling its impact on the system. In contrast, \cite{9839429} introduces an "always-ON" protocol, where the RIS remain active throughout the channel estimation process. This not only eliminates the need for on-off amplitude control but also offers a more realistic implementation.

Recently, research has expanded to multi-RIS-assisted systems, which introduce additional challenges and opportunities in channel estimation due to increased signal reflections and interference. These systems require novel estimation strategies to handle more complex channel models and optimize the coordination between multiple RIS units. Furthermore, studies have explored the impact of RIS placement \cite{10439018,10558715}, efficient RIS selection \cite{10623725}, integrated sensing and communications \cite{10497119}, and channel estimation \cite{10767769}, in order to develop robust and scalable solutions that improve the efficiency and reliability of multi-RIS-assisted wireless networks.


Motivated by the aforementioned factors, our work considers the uplink transmission of a coded passive multi-RIS-assisted MIMO system.  To the best of our knowledge, this is the first study to employ iterative coding techniques to enhance the channel estimation in RIS-assisted MIMO systems. Specifically, we propose an iterative detection, decoding, and estimation scheme that encodes pilots and exploits them and the parity bits to perform an enhanced channel estimation. We also develop an efficient iterative channel estimator for multi-RIS-assisted systems, which accounts for both LOS and NLOS scenarios, while reducing pilot overhead and exploiting the benefits of the coding scheme. Numerical results show that the proposed scheme and channel estimator outperforms competing techniques in LOS and NLOS scenarios.

The remainder of this paper is structured as follows. Section II presents the proposed system model. In Section III, the proposed iterative channel estimation technique is outlined. Section IV discusses the simulation results and Section V concludes the paper.

\textit{Notation:} Bold capital letters represent matrices, while bold lowercase letters denote vectors. The symbol $\mathbf{I_n}$ refers to an $n \times n$ identity matrix and $\text{diag}(\mathbf{A})$ is a diagonal matrix with the diagonal elements of $\mathbf{A}$. The sets of complex and real numbers are denoted by $\mathbb{C}$ and $\mathbb{R}$, respectively. [·]$^{-1}$, [·]$^{T}$, and [·]$^H$ denote the inverse, transpose, and conjugate transpose, respectively. The Kronecker product is represented by $\otimes$.

\begin{figure*}
    \centerline{\includegraphics[width=0.700\textwidth]{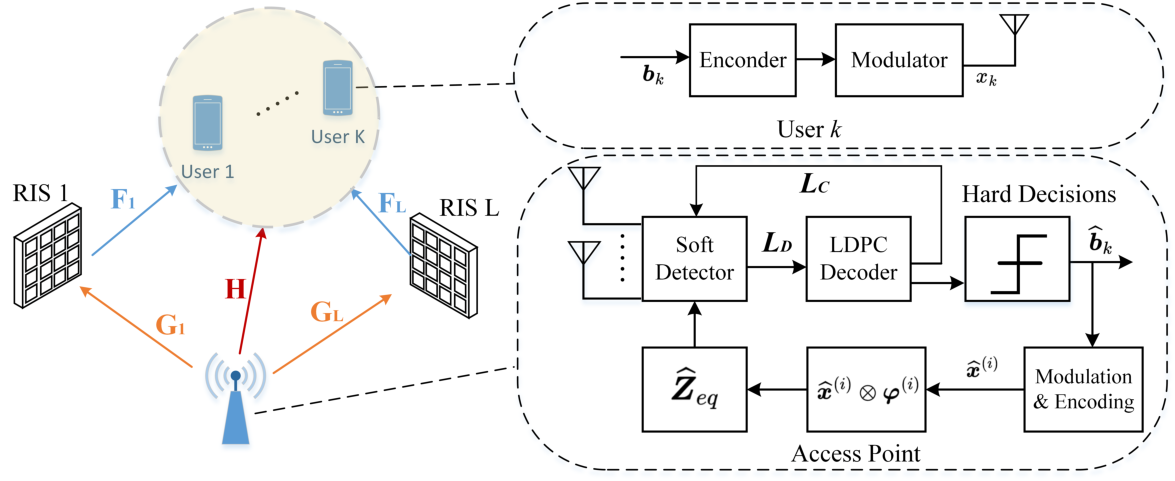}}
    \vspace{-0.75em}
    \caption{System model of an IDD multiuser multiple-antenna system.}
    \label{fig:blockdiagram}
    \vspace{-1.2em}
\end{figure*}
   \section{System Model}

A single-cell multiuser system featuring a large-scale multiple-antenna access point (AP) \cite{mmimo,wence} assisted by $L$ reflective intelligent surfaces (RIS) is considered, as illustrated in Fig. \ref{fig:blockdiagram}. In this configuration, the AP is equipped with $M$ antennas that support $K$ users, each equipped with a single antenna. The information bits for each user are encoded via individual LDPC channel encoders and subsequently modulated to $x_k$ using a QPSK modulation scheme. The transmit symbols $x_k$ have zero mean and share the same energy, with $E[|x_k|^2] = \sigma^2_x$. These modulated symbols are transmitted over block-fading channels.

Each RIS has $N$ reflecting elements, and their reflection coefficients are modeled as a complex vector $\boldsymbol{\varphi}_j \triangleq [e^{j\theta_1}, \dots, e^{j\theta_N}]^T$, where $\theta_n$ represents the phase shift of the $n$th unit of the RIS-$j$. The corresponding diagonal reflection matrix at time instant $i$ is given by and $\mathbf{\Phi}_j^{(i)} \triangleq \text{diag}(\boldsymbol{\varphi}_j)$. The received signal $\mathbf{y}^{(i)}$ at time instant $i$ can be expressed as

\begin{equation}
    \boldsymbol{y}^{(i)} = (\mathbf{H} + 
    \sum_{j=1}^L\mathbf{G}_j\mathbf{\Phi}_j^{(i)}\mathbf{F}_j +
     \mathbf{E}^{(i)})
    \boldsymbol{x}^{(i)} + \mathbf{n}^{(i)},
    \label{eq01}
\end{equation}
where $\mathbf{H} \in \mathbb{C}^{M \times K}$, $\mathbf{G}_j \in \mathbb{C}^{M \times N}$, and $\mathbf{F}_j \in \mathbb{C}^{N \times K}$ represent the communication links from the AP to the users, from the AP to RIS-$j$, and from RIS-$j$ to the users, respectively. The vector $\mathbf {x}^{(i)} \triangleq [x_1^{(i)}, \dots, x_K^{(i)}]^T$ represents the coded symbols transmitted by the users at time instant $i$, while and $\mathbf{n}^{(i)} \sim \mathcal{CN}(\mathbf{0}_M, \sigma_n^2\mathbf{I}_M)$ represents the noise. 

The signal reflected between different RIS at instant $i$ is represented by $\mathbf{E}^{(i)}$. Due to the multiplicative fading effect \cite{9998527}, this element has a negligible effect on the received signal and is ignored in the remainder of this work. Note that RIS can also be deployed to reduce the effect of $\mathbf{E}^{(i)}$.

The contributions of different RIS to the users can be grouped in a concise expression by
\begin{align}
\boldsymbol{y}^{(i)} =&  [\mathbf{G}_1 \dots \mathbf{G}_L]
\begin{bmatrix}
\boldsymbol{\Phi}_1^{(i)} & \boldsymbol{0} & \boldsymbol{0} \\
\boldsymbol{0} & \ddots           & \boldsymbol{0} \\
\boldsymbol{0} & \boldsymbol{0}   & \boldsymbol{\Phi}_L^{(i)} \\
\end{bmatrix} 
\begin{bmatrix}
\mathbf{F}_1 \\
\vdots \\
\mathbf{F}_L
\end{bmatrix}\boldsymbol{x}^{(i)}
\\
&+ \mathbf{H}\boldsymbol{x}^{(i)} + \boldsymbol{n}^{(i)}. \nonumber 
\end{align}
Grouping the matrices of the communication links leads to
\begin{equation}
    \boldsymbol{y}^{(i)} = (\mathbf{H} + \mathbf{G}_p\boldsymbol{\Phi}_p^{(i)}\mathbf{F}_p)\boldsymbol{x}^{(i)} + \boldsymbol{n}^{(i)} = \mathbf{\bar{H}}_p^{(i)}\boldsymbol{x}^{(i)}  + \boldsymbol{n}^{(i)}, 
    \label{eq03}
\end{equation}
where $\boldsymbol{\bar{H}}_p$ represents the equivalent channel between the AP and the users.

Analyzing (\ref{eq03}), this equation is similar to that used for the representation of a single RIS-assisted MIMO system, which differs only in how the matrices are grouped. Therefore, the same techniques can be used here. In addition to the grouped RIS phase-shift matrix, $\boldsymbol{\Phi}_p^{(i)}$ remains a diagonal-only matrix. Since the matrix $\mathbf{\Phi}$ is diagonal, the received signal in (\ref{eq03}) can also be written in terms of $\boldsymbol{\varphi}^{(i)}_p$ as given by 
\begin{equation}
     \boldsymbol{y}^{(i)} = \sum_{k=1}^K\boldsymbol{h}_kx_k^{(i)} + \sum_{k=1}^K\mathbf{Z}_k\boldsymbol{\varphi}_p^{(i)}x_k^{(i)} + 
     \mathbf{n}^{(i)},
    \label{eq:eq04}
\end{equation}
where $\mathbf{Z}_k = \mathbf{G}_p \text{diag}(\boldsymbol{f}_{p,k})$, with  $\boldsymbol{f}_{p,k}$  denoting the 
$k$th column of the matrix $\mathbf{F}_p$.

An estimate $\hat{x}_k$ of the transmitted symbol is obtained by applying a linear receive filter $\boldsymbol{w_k}$ to the received signal:
\begin{equation}
    \hat{x}_k^{(i)} = (\mathbf{w}_k^{(i)})^H\boldsymbol{y}^{(i)}.
    \label{detection_estimate_1}
\end{equation}

\subsection{Enhancing Detection through SIC}
\label{subsec:SIC}
In the SIC detector \cite{spa,mfsic,dfcc,mbdf,bfidd,detmtc,msgamp1,msgamp2,1bitidd,dynovs,comp,c&d_idd,idd_llr,idd_ocl}, the received vector $\boldsymbol{y}^{(i)}$ is processed by demapping, where a log-likelihood ratio (LLR) is calculated for each bit in the transmit vector $\boldsymbol{x}^{(i)}$. For simplicity, we omit the $(i)$ notation in this section. Therefore, the extrinsic LLR value $L_D$ for the $\upsilon$th code bit $b_\upsilon$ is computed as follows:
\begin{equation}
    L_{D}(b_{\upsilon}) = \log \frac{\sum\nolimits_{\boldsymbol{x} \in \mathcal{X}_{\upsilon}^{+1}} P(\boldsymbol{y} \vert \boldsymbol{x}, \mathbf{\bar{H}}_p) P(\boldsymbol{x})} {\sum\nolimits_{\boldsymbol{x} \in \mathcal{X}_{\upsilon}^{-1}} P(\boldsymbol{y} \vert \boldsymbol{x}, \mathbf{\bar{H}}_p) P(\boldsymbol{x})} - L_{C}(b_{\upsilon}).
    \label{eq:ldlc}
\end{equation}

Inspired by prior work on IDD schemes \cite{spa,mfsic,mbdf,8240730}, the soft estimate of the $k$th transmitted symbol is firstly calculated based on the $\boldsymbol{L}_C$ (extrinsic LLR) provided by the channel decoder from a previous stage:
\begin{equation*} 
\tilde {x}_{k}=\sum _{x\in \mathcal {A}}x\text {Pr}(x_{k}=x)=\sum _{x\in \mathcal {A}}x\left ({\prod _{l=1}^{M_{c}}\left [{1+\text {exp}(-x^{l}L_{c}^{l})}\right]^{-1}}\right), 
\end{equation*}
where $\mathcal {A}$ is the complex constellation set with $2^{M_c}$ possible points. The symbol $x^l$ corresponds to the value $(+1, -1)$ of the $l$th bit of the symbol $x$. 

A symbol estimate uses SIC, where the value of $\boldsymbol{\varphi}$ is fixed and $\mathbf{w_k}$ is chosen to minimize the mean square error (MSE) between the transmitted symbol $x_k$ and the filter output:
{\begin{equation} 
    \mathbf {w}_{k}=\arg \min _{ \tilde{\mathbf{w}}_{k}} E\left [{\left \vert{ x_{k}-\tilde{\mathbf{w}}_{k}^{H}\mathbf {y}_k}\right \vert ^{2}}\right]. 
\end{equation}}
It can be shown that the solution is given by
\begin{equation}
    \mathbf{w}_k = \left(\frac{\sigma^2_n}{\sigma^2_x}\mathbf{I_{n_r}} + \mathbf{\bar{H}}_p \mathbf{\Delta}_k\mathbf{\bar{H}}_p^{\rm H} \right)^{-1}\boldsymbol{\bar{h}}_k,
    \label{eq:w}
\end{equation}
where $\mathbf {\bar{H}} \triangleq [\mathbf{\bar{h}_1}, \dots, \mathbf{\bar{h}_K}]^H$ is the equivalent channel and the covariance matrix $\mathbf{\Delta_k}$  is
{\begin{equation}
    \mathbf{\Delta}_k = \text{diag}\left[\frac{\sigma^2_{x_{1}}}{\sigma^2_x}\dots \frac{\sigma^2_{x_{k-1}}}{\sigma^2_x}, 1, \frac{\sigma^2_{x_{k+1}}}{\sigma^2_x},\dots,\frac{\sigma^2_{x^2_{K}}}{\sigma^2_x}  \right],
\end{equation}}
where $\sigma^2_{x_{i}}$ is the variance of the $k$th user, computed as:
{\begin{equation}
\sigma_{x_{k}}^{2}=\sum\limits_{x\in {\cal A}}\vert x-\bar{k} _{i}\vert ^{2}P(x_{k}=x). 
\end{equation}}

\subsection{Design of Reflection Parameters}

The optimization of reflection parameters builds upon the work in \cite{10747209}, which focuses on the design of reflection parameters using the minimum mean square error (MMSE) criterion. This approach yields effective results for refining LLRs within an IDD system. Specifically, an MMSE receiver filter was employed to facilitate SIC at the receiver. The optimization of the reflection parameters is given by
\begin{equation} 
\boldsymbol{\varphi}_o = \boldsymbol{\beta}^{-1}\boldsymbol{\Psi},
\label{eq05}
\end{equation}
where
\begin{equation}
    \boldsymbol{\beta} = \sum^{K}_{k=1}(\mathbf{WZ}_k)^H(\mathbf{WZ}_k),
\end{equation}
\begin{equation}
    \boldsymbol{\Psi} =\sum_{k=1}^K(\mathbf{WZ}_k)^H)(\boldsymbol{e}_k-\mathbf{W}\boldsymbol{\bar{h}}_k),
\end{equation}
$\boldsymbol{e}_k$ is a column vector with zeros, except for the one in the $k$th element \cite{jidf}. This method involves truncating the reflection parameters to satisfy the passive RIS constraint  $|[\boldsymbol{\varphi}]_n|=1$, for $\forall n$, which leads to following truncation: 
\begin{equation}
    [\boldsymbol{\varphi_t}^\text{passive}]_i = \frac{[{\varphi}_o]_i}{|[{\varphi}_o]_i|}.
\end{equation}
    \section{Proposed Channel Estimation}
\begin{figure}
\centerline{\includegraphics[width=0.5\textwidth]{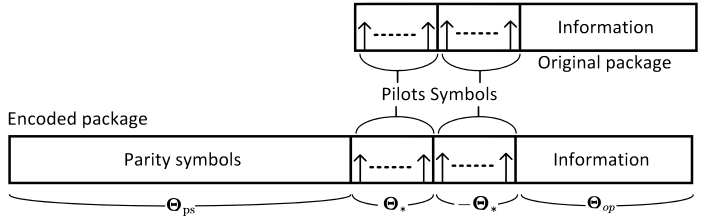}}
    \vspace{-0.725em}
    \caption{Systematically encoded packet and original post-modulation packet.}
    \label{fig:package}
    \vspace{-1em}
\end{figure}

In this section, we propose a novel channel estimation technique for multi-RIS-assisted MIMO systems that leverages iterative detection and decoding with EP and LDPC decoding to reduce the number of pilots required to estimate the cascaded RIS coefficients. All the pilots symbols are encoded with the data bits using a systematic encoder; therefore, the pilots remain unaltered and can be considered known by the receiver, as illustrated in Fig. \ref{fig:package}. 
The technique begins by estimating the direct channel and obtaining an initial raw estimation of the reflected RIS channel. After the first iteration, the entire estimated symbol packet is used as pilots to refine the estimation of the reflected channel. Since the receiver is prone to incorrect symbols estimates, this procedure is performed iteratively to refine the channel estimation in each iteration.

To introduce this approach, we first explain how the direct channel can be estimated, then describe the coarse estimation of the reflected channel, and eventually show how EP and the iterative processing enhance the overall estimation accuracy.

\subsection{Direct Channel Estimation}
To estimate the direct channel, we used an always-on channel estimation approach without switching off the selected RIS elements \cite{9839429}. This protocol involves dividing the pilots sequence into two partitions of equal size, allowing us to represent the received signal for each partition as

\begin{equation}
    \boldsymbol{y}^{(j)} = (\mathbf{H} + \mathbf{G}_p\text{diag}(\boldsymbol{\varphi}_p^{(j)})\mathbf{F}_p)\boldsymbol{x}^{(j)} + \boldsymbol{n}^{(j)},
    \label{eq:partA}
\end{equation}
where $j$ belongs to the set of pilot symbols. We define the first and second partitions, respectively, as:  
\begin{equation}
\mathcal{P}_1 = \left\{ j \mid t \leq j \leq t + \frac{N_p}{2}-1\right\}    
\label{eq:partition01}
\end{equation}
\begin{equation}
\mathcal{P}_2 = \left\{ j \mid t + \frac{N_p}{2} \leq j \leq t + N_p -1\right\}.    
\label{eq:partition02}
\end{equation}

By selecting the pilots and reflection parameter values such that $x^{(j)} = x^{\left(j + \frac{N_p}{2}\right)}$  and $\boldsymbol{\varphi}_p^{(j)} = -\boldsymbol{\varphi}_p^{\left(j + \frac{N_p}{2}\right)}$, we can define the sum and subtraction of each received signal as

\begin{equation}
    \frac{\boldsymbol{y}^{(j)}+\boldsymbol{y}^{(j+\frac{Np}{2})}}{2} = \mathbf{H}\boldsymbol{x}^{(j)} + \boldsymbol{w}
    \label{eq:directchannel}
\end{equation}
and
\begin{equation}
    \frac{\boldsymbol{y}^{(j)}-\boldsymbol{y}^{(j+\frac{Np}{2})}}{2} = \textbf{G}_p\text{diag}(\boldsymbol{\varphi}_p^{(j)})\textbf{F}_p\boldsymbol{x}^{(j)} + \boldsymbol{w},
    \label{eq:reflectedchannel}
\end{equation}
where $\boldsymbol{w} \sim \mathcal{CN}(\mathbf{0_M}, \frac{\sigma_n^2}{2} \mathbf{I_M})$.

From (\ref{eq:directchannel}), we can apply conventional channel estimation methods to estimate only the direct channel. In this work, we use the linear minimum mean square error (LMMSE) channel estimator, given by 
\begin{equation}
    \hat{\mathbf{H}}_\text{LMMSE} = \mathbf{Y}_p\left(\mathbf{P}^H\mathbf{R}_\text{H}\mathbf{P}+\frac{\sigma_n^2}{2\sigma_x^2}\mathbf{I}\right)^{-1}\mathbf{P}^H\mathbf{R}_\text{H},
    \label{eq:lmmse}
\end{equation}
where $\mathbf{R}_\text{H} := E[\mathbf{H}\mathbf{H}^H]$ is the channel covariance matrix, $\mathbf{P}$ is the matrix representing the vectors of pilot symbols, and  $\mathbf{Y}_p$ is the received matrix obtained by summing the received signals from (\ref{eq:directchannel}). For this estimation, since both partitions are used to obtain the direct channel, a total of  $N_p = 2K$ pilot symbols are required.

\subsection{Cascate RIS Channel Estimation}
To estimate the coefficients of the reflected channel without interference from the direct link, we use (\ref{eq:reflectedchannel}). By rewriting the received signal in terms of $\boldsymbol{\varphi}_p^{(i)}$ using (\ref{eq:eq04}), the received signal from the RIS can be expressed as

\begin{equation}
    \boldsymbol{y}^{(j)}_\text{cascate}=\frac{\boldsymbol{y}^{(j)}-\boldsymbol{y}^{(j+\frac{Np}{2})}}{2} = \sum_{k=1}^K\mathbf{Z}_k\boldsymbol{\varphi}_p^{(j)}x_k^{(j)} + \boldsymbol{w}^{(j)}.
    \label{eq:cascate_1}
\end{equation}

Eq. \eqref{eq:cascate_1} can be rearranged by concatenating the matrices $\mathbf{Z}_k$ and eliminating the summation, resulting in 

\begin{align}
    \boldsymbol{y}^{(j)}_\text{cascate}&= [\mathbf{Z}_1, \dots, \mathbf{Z}_K](\boldsymbol{x}^{(j)}\otimes{\varphi}_p^{(j)} ) + \boldsymbol{w}^{(j)}
    \\
    &= \mathbf{Z_\text{all}}\boldsymbol{\lambda}^{(j)} + \boldsymbol{w}^{(j)} \nonumber 
\end{align}
where $\mathbf{Z_\text{all}} \in \mathbb{C}^{M \times KN}$ is the concatenated matrix of the cascaded channel for each user and $\boldsymbol{\lambda}^{(j)}=(\boldsymbol{x}^{(j)}\otimes{\varphi}_p^{(j)} )$ is a complex vector with KN elements.

Since the symbols of partition $\mathcal{P}_1$ are known by the receiver, we can assume that $\boldsymbol{\lambda}^{(j)}$ is also fully known. After $T$ time slots of pilot transmission, we can obtain the $M \times T$ overall measurement matrix $\mathbf{Y}_\text{cascate}=[\boldsymbol{y}_\text{cascate}^{(1)}, \dots,\boldsymbol{y}_\text{cascate}^{(T)}]$ as 
\begin{equation}
    \mathbf{Y}_\text{cascate}= 
    \mathbf{Z_\text{all}}\boldsymbol{\Lambda} + \mathbf{W},  
    \label{eq:cascate_2}
\end{equation}
where $\boldsymbol{\Lambda} = [\boldsymbol{\lambda}^{(1)}, \dots, \boldsymbol{\lambda}^{(T)}]$ and $\mathbf{W} = [\boldsymbol{w}^{(1)}, \dots, \boldsymbol{w}^{(T)}]$.

This expression can be considered equivalent to a conventional MIMO channel estimation problem, which allows us to apply (\ref{eq:lmmse}) to estimate the coefficients. 

To mitigate multiuser interference, the matrix $\boldsymbol{\Lambda}$ should be orthogonal or semi-orthogonal, ensuring that the coefficients of $\boldsymbol{Z_\text{all}}$ can be estimated independently for each user. This matrix should be well conditioned to prevent noise amplification during equation inversion. To this end, we generate $\boldsymbol{\lambda}^{(i)}$ using Hadamard sequences for $\boldsymbol{x}^{(i)}$, while $\boldsymbol{\varphi}_p^{(i)}$ is derived from the DFT matrix \cite{9130088}. Note that to estimate all channel coefficients, $\boldsymbol{\Lambda}$ must have at least a rank of $KN$, which may be infeasible in some scenarios. Therefore, we suggest first obtaining a coarse approximation of the cascaded RIS channel using only a few pilots (number of pilots $\ll KN$), and then refining the estimates through iterative channel estimation.

\subsection{Proposed Iterative Channel Estimation}

In this step, our goal is to use the entire decoded symbol sequence as input, even if some symbols are incorrectly decoded. As shown in Fig. \ref{fig:blockdiagram}, the output of the IDD provides the decoded bits. We employ the encoded pilot scheme, which combines pilot bits with information bits using a systematic encoder. This enables iterative channel estimation and decoding, where the EPs contribute to both processes, enhancing overall performance \cite{4357052}. For iterative channel estimation, the first step is to reapply coding and modulation to transform the decoded bits back into symbols. Then, we compute the Kronecker product $(\boldsymbol{x}^{(j)}\otimes{\varphi}_p^{(j)} )$ to derive $\boldsymbol{\Lambda}$, which allows the application of standard MIMO channel estimation techniques to (\ref{eq:cascate_2}). However, it is crucial for $\boldsymbol{\Lambda}$ to remain semi-orthogonal and well-conditioned. Since the phase configurations that satisfy these conditions differ from those obtained using (\ref{eq05}), we adopt a non-optimal approach for the parity bits.

Assuming that the package in Fig. \ref{fig:package} consists of $N_\mathrm{ps}$ parity symbols, $N_\mathrm{p}$ pilot symbols and $N_\mathrm{info}$ information symbols, and that $\boldsymbol{\Theta}_\mathrm{i} = [\boldsymbol{\varphi}^{(i)} \boldsymbol{\varphi}^{(i+1)} \dots]$ represents the concatenation of the reflection parameter vectors, we can express the reflection parameter vectors for each symbol as follows:
\begin{equation}
    \begin{matrix}  
    {\boldsymbol{\Theta}}_{\mathrm{ps}} =\left [{
\begin{array}{cccc} 
1 &1  &\cdots &1 \\ 
1 &\omega  &\cdots &\omega ^{N_\mathrm{ps}-1} \\ 
\vdots  &\vdots &\ddots &\vdots \\ 
1 &\omega ^{N-1}  &\cdots &\omega ^{(N-1)(N_\mathrm{ps}-1)} \end{array}}\right]\end{matrix},
\end{equation}
\begin{equation}
    \boldsymbol{\Theta}_\mathrm{p} = [\boldsymbol{\Theta}_\mathrm{*} -\boldsymbol{\Theta}_\mathrm{*}],
\end{equation}
\begin{equation}
    \begin{matrix}  
    {\boldsymbol{\Theta}}_{*} =\left [{
\begin{array}{cccc} 
1 &1  &\cdots &1 \\ 
1 &\varpi  &\cdots &\varpi ^{N_\mathrm{p}/2-1} \\ 
\vdots  &\vdots &\ddots &\vdots \\ 
1 &\varpi ^{N-1}  &\cdots &\varpi ^{(N-1)(N_\mathrm{p}/2-1)} \end{array}}\right]\end{matrix},
\end{equation}

\begin{equation}
    \boldsymbol{\Theta}_{\mathrm{o}} = [\boldsymbol{\varphi}_\mathrm{o} 
    \dots
    \boldsymbol{\varphi}_\mathrm{o}], \quad
    \boldsymbol{\varphi}_\mathrm{o} \text{ computed with (\ref{eq05})}
\end{equation}
where $\omega = e^\frac{-2\pi i}{N_\mathrm{ps}}$ and $\varpi = e^\frac{-4\pi i}{N_\mathrm{p}}$. If $N<\frac{N_\mathrm{p}}{2}$, to ensure pseudo-orthogonality between the RIS elements, we concatenate $N \times N$ DFT matrices such that the condition $uN \geq \frac{N_\mathrm{p}}{2}$ is met, where $u \in \mathbb{N}^*$ is the number of concatenated matrices. 
This ensures pseudo-orthogonality in $\boldsymbol{\Lambda}$ for both sequences of parity bits and pilots, while preserving the information symbols with the optimal reflection parameters. Note that since the same optimal phase configuration is applied to the information symbols, $\boldsymbol{\Theta}_{\mathrm{o}}$ is a low-rank matrix, which minimally contributes to the channel estimation.


The initial symbol estimates may contain errors, which can propagate to the channel estimation. To address this issue, the channel estimates are iteratively refined and the symbols are reprocessed until convergence is achieved or a predefined maximum number of iterations is reached. The pseudo-code of the proposed algorithm is shown below.

\begin{algorithm}[H]
\footnotesize
    \label{algor1}
    \begin{algorithmic}[1]
        \caption{Proposed Iterative Channel Estimation}\label{alg:cap}
        \STATE Estimate the direct channel ${\mathbf{H}}$ using (\ref{eq:lmmse}) based on (\ref{eq:directchannel}).
        \STATE Perform initial estimation of the reflected channel ${\boldsymbol{Z_\text{all}}}$ using (\ref{eq:lmmse}) based on (\ref{eq:reflectedchannel}).
        \FOR{$t = 1$ to Max Iterations or until $\text{tol} > \text{NMSE}$}
            \STATE \textbf{IDD Scheme - SIC}: Apply the scheme as defined in Sec. \ref{subsec:SIC}.
            \STATE Refine the reflected channel estimation $\hat{\boldsymbol{Z}}_\text{all}$ using (\ref{eq:cascate_2}) and the estimator (\ref{eq:lmmse}).
        \ENDFOR
    \end{algorithmic}
\end{algorithm}
\vspace{-0.5cm}

     \section{Numerical Results}

A short-length regular LDPC code \cite{memd,vfap} with a block length of $N=512$ and rate $R = 1/2$ was considered with QPSK modulation. The channel is assumed to experience block fading and the estimation of the channel state information (CSI) is done at the receiver (AP). Two uplink scenarios were evaluated: LOS ( weakened direct link with strong reflected link) and NLOS (no direct link and strong reflected link). The path loss models follow the 3GPP standard \cite{access2010further}, and the system parameters are presented in Table \ref{tab:parameters}. To account for small-scale fading effects, a Rayleigh fading channel model was adopted for all channels. \vspace{-0.75em}

\begin{table}[ht]
    \caption{Simulation Parameters}\vspace{-0.75em}
    \label{tab:parameters}
\centering
\begin{tabular}{c|c}
\hline
{\textbf{Parameters}}               & { \textbf{Values}} \\ \hline
Frequency                           & 5 GHz              \\ \hline
Bandwidth                           & 1 MHz              \\ \hline
Noise power spectral density        & -170 dBm/Hz              \\ \hline
Path loss AP-RIS; RIS-Users (dB)    & $37.3+22log_{10}(d)$  \\ \hline
Path loss AP-Users (dB)             & $32.4+30log_{10}(d)$  \\ \hline
Numer of AP antennas                & 8       \\ \hline
Number of Users                     & 4       \\ \hline
Number of RISs                      & 2      \\ \hline
Number of Cells (per RIS)           & 16     \\ \hline
Location of AP                      & (0 m, 0 m, 0 m)       \\ \hline
Location of RIS$_1$                 & (500 m, 10 m, 0 m)     \\ \hline
Location of RIS$_2$                 & (500 m, -10 m, 0 m)    \\ \hline
Geometric center of users positions & (500 m, 0 m, 0 m)      \\ \hline
Users Spatial Radius                & 5 m                   \\ \hline
\end{tabular}
\end{table}

\begin{figure}
    \vspace{-1em}
\includegraphics[width=0.45\textwidth,height=5.0cm]{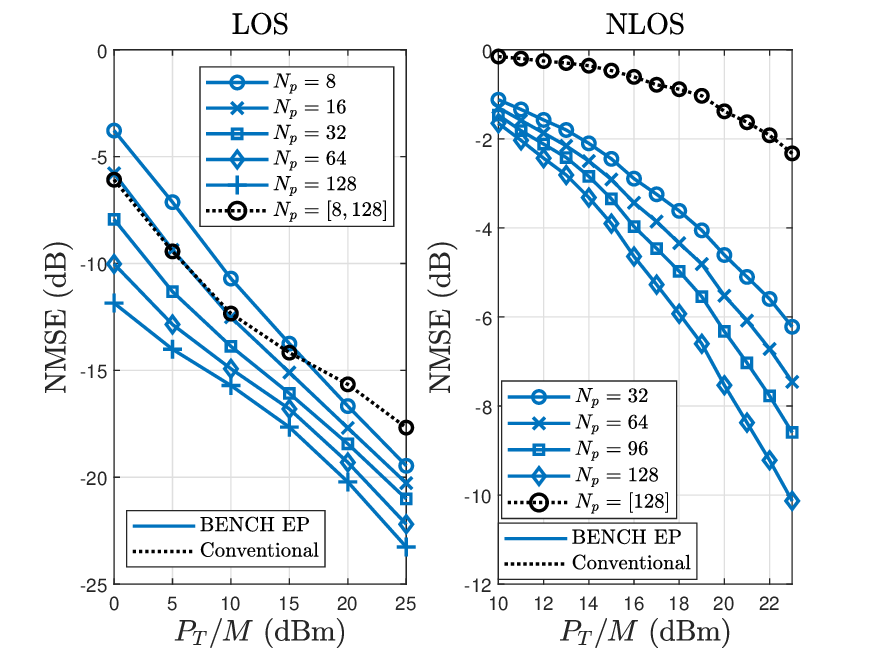}    \vspace{-1.0em}
    \caption{Comparison between the proposed and conventional on-off channel estimation techniques.}
    \label{fig:spawc_00}
\vspace{-0.5em}
\end{figure}

Fig. \ref{fig:spawc_00} presents the NMSE performance comparison as a function of the users' transmission power $P_t$. The figure compares the proposed technique with the conventional estimation approach, which employs an on-off scheme where pilots are used separately for direct estimation and the reflected link.

In the LOS scenario, the proposed technique achieves the same estimation accuracy using only $N_p=16$ pilots at a low SNR, compared to 136 pilots required by the conventional approach (8 pilots for direct link estimation and 128 for the reflected link). In the NLOS scenario, the proposed technique requires a higher number of pilot symbols due to the absence of a direct link. However, it still achieves a significant performance gain compared with the conventional system.

\begin{figure}
    \vspace{-1em}
\includegraphics[width=0.45\textwidth,height=5.0cm]{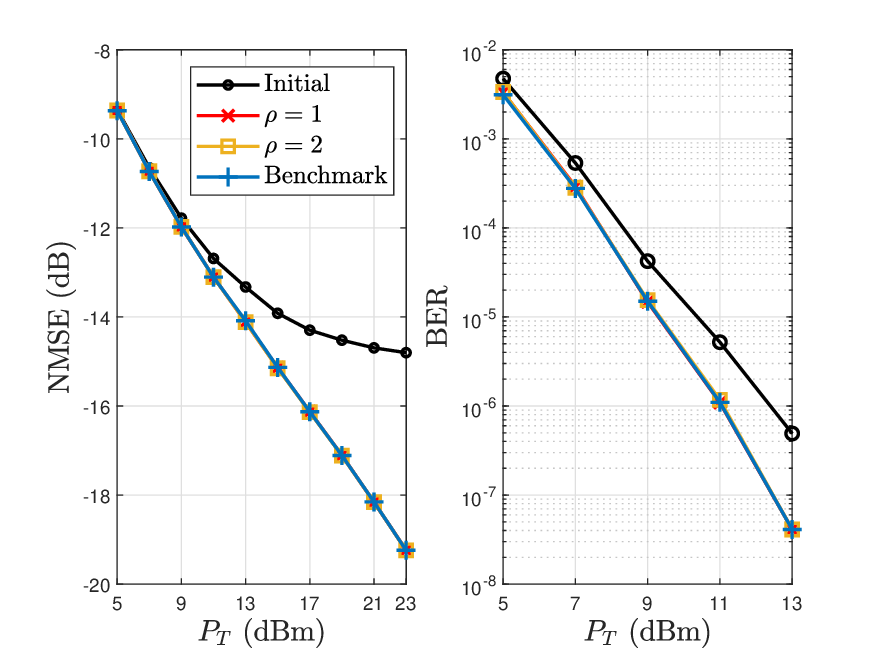}
  \vspace{-1em}
    \caption{LOS RIS-Assisted system with $N_p=16$.}
    \label{fig:spawc_01}
    \vspace{-0.5em}
\end{figure}

\begin{figure}
    \vspace{-1em}
\includegraphics[width=0.45\textwidth,height=5.0cm]{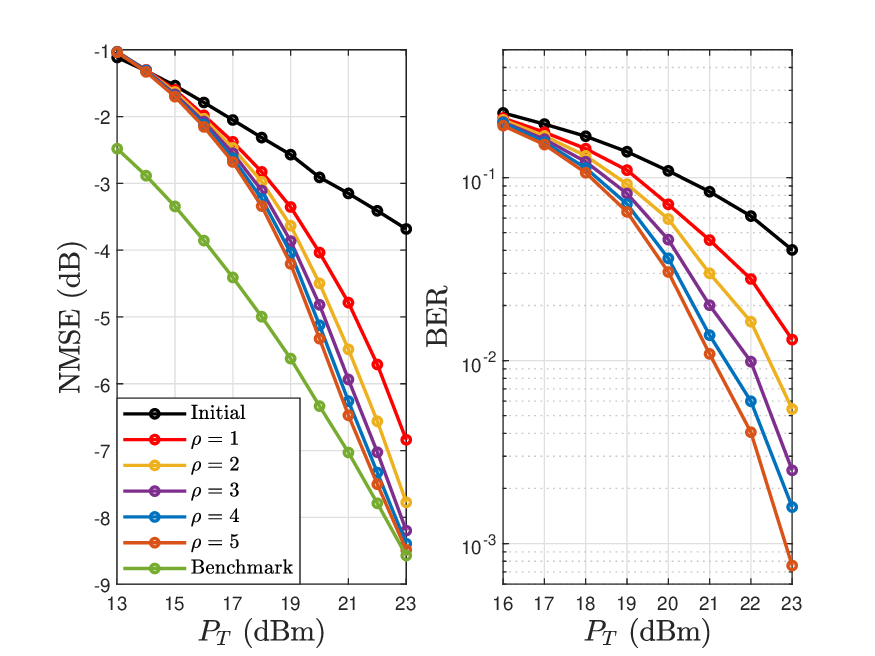}
  \vspace{-1em}
    \caption{NLOS RIS-Assisted system with $N_p=96$.}
    \label{fig:spawc_02}
    \vspace{-2em}
\end{figure}

The derivation of an analytical expression for the optimal number of pilots is challenging, as it depends on factors such as channel models, LDPC codes and code rates \cite{4357052}. The results in terms of the NMSE and BER as functions of $P_T$ for two specific scenarios, LOS and NLOS, are depicted in Figs. \ref{fig:spawc_01} and \ref{fig:spawc_02}, respectively, where $\rho$ denotes the number of iterations in the channel estimation refinement procedure. An improvement of performance is obtained as the number of pilots increases. In the LOS case, the benchmark performance is achieved with only a few iterations ($\rho=1$) due to the presence of a direct link, which significantly aids in estimating the received symbols. Conversely, in the NLOS scenario, in which the direct link is absent, multiple iterations (depending on the SNR) are required to reach the benchmark. In both cases, NMSE and BER improvements were observed, particularly at medium-to-high SNR levels. \vspace{-0.5em}

    \section{Conclusion}
In this study, we proposed a novel iterative detection, decoding, and channel estimation scheme for RIS-assisted MIMO systems. Unlike existing approaches, our method exploits coding in the uplink to use parity bits for both decoding and channel estimation while employing encoded pilots to enhance performance. The proposed scheme significantly reduces the minimum number of pilots required for both LOS and NLOS scenarios, achieving superior performance in LOS scenarios. Numerical results show that our approach has large performance gains in terms of channel estimation NMSE and BER. \vspace{-0.25em}

    \bibliography{support/main}
\end{document}